\documentclass[conference]{IEEEtran}
\makeatletter
\def\ps@headings{%
\def\@oddhead{\mbox{}\scriptsize\rightmark \hfil \thepage}%
\def\@evenhead{\scriptsize\thepage \hfil \leftmark\mbox{}}%
\def\@oddfoot{}%
\def\@evenfoot{}}
\makeatother

\usepackage{amsmath}
\usepackage{amssymb}
\usepackage{bbm}
\usepackage{graphics}
\usepackage{psfrag}
\usepackage{epsfig}
\usepackage[verbose]{cite}
\usepackage{algorithm}
\usepackage{algorithmic}
\usepackage{cite}
\newcommand{\ora}{\overrightarrow}
\newcommand{\G}{\ora{G}}
\newcommand{\E}{\ora{E}}
\newcommand{\e}{\ora{e}}
\newcommand{\ve}{\varepsilon}
\newcommand{\vt}{\vartheta}

\newcommand{\mcc}{\mathcal{C}}

\newcommand{\ol}{\overline}

\newcommand{\mc}{\mathcal}

\newtheorem{prop}{Proposition}

\title{Throughput and Latency of Acyclic Erasure Networks with Feedback in a Finite Buffer Regime}
\author{Nima Torabkhani$^\dag$,
     Badri N. Vellambi$^{\dag\ddag}$
     Faramarz Fekri$^\dag$\\
     $^\dag$ School of Electrical and Computer Engineering,
    Georgia Institute of Technology\\
     $^{\dag\ddag}$ Institute for Telecommunications Research,
    University of South Australia\\
     E-mail: \{nima, fekri\}@ece.gatech.edu, badri.vellambi@unisa.edu.au}

\begin{document}
\maketitle

\begin{abstract}
The exact Markov modeling analysis of erasure networks with finite buffers is an extremely hard problem due to the large number of states in the system. In such networks, packets are lost due to either link erasures or blocking by the full buffers. In this paper, we propose a novel method that iteratively estimates the performance parameters of the network and more importantly reduces the computational complexity compared to the exact analysis. This is the first work that analytically studies the effect of finite memory on the throughput and latency in general wired acyclic networks with erasure links. As a case study, a random packet routing scheme with ideal feedback on the links is used. The proposed framework yields a fairly accurate estimate of the probability distribution of buffer occupancies at the intermediate nodes using which we can not only identify the congested and starving nodes but also obtain analytical expressions for throughput and average delay of a packet in the network. The theoretical framework presented here can be applied to many wired networks, from Internet to more futuristic applications such as networks-on-chip under various communication and network coding scenarios.

\end{abstract}
\section{Introduction}\label{FB-intro}
In networks, packets may have to be relayed through a series of intermediate nodes where each may receive packets via many other data streams as well. Hence, the packets may have to be stored at intermediate nodes for transmission at a later time. For infinte buffer case, the intermediate nodes need not have to drop the arriving packets. However, often times, buffers are limited in size. Although a large buffer size is usually affordable and preferred to minimize packet drops, large buffers have an adverse effect on the packet delay. Additionally, as second-order effects, using larger buffer sizes at intermediate nodes would have practical problems such as on-chip board space and increased memory-access latency.

The problem of buffer sizing and congestion control is of paramount interest to router design engineers. Typical routers today route several tens of gigabits of data each second \cite{FlowRef}. Realistic studies have shown that, at times, Internet routers handle about ten thousand independent streams/flows of data packets. With a reasonable buffer size of few Gigabytes of data, each stream can only be allocated a few tens of data packets. Therefore, at times when long parallel flows congest a router, the effects of such a small buffer space per flow come to play. Though we are motivated partly by such concerns, our work is far from modeling realistic scenarios. This work modestly aims at providing a theoretical framework to understand the fundamental limits of single information flow in finite-buffer wired networks and investigate the trade-offs between throughput, average packet delay and buffer size.

The problem of computing capacity and designing efficient coding schemes for lossy networks has been widely studied \cite{AmirDana:capacity, PakzadF05, YeungNCJrnl}. However, the study of capacity of networks with finite buffer sizes has been limited. This can be attributed solely to the fact that finite buffer systems are analytically harder to track in general. In \cite{VelambiITW, VelambiITA}, it was shown that min-cut capacity cannot be achieved due to the limited buffer constraint in a line network. Here, we wish to establish a framework to investigate the same for general wired networks.

Recently, \cite{PakzadF05} outlined various coding strategies for achieving capacity in line erasure networks. However, it only considered infinite buffer in its study. Later, \cite{NetCod:Lun} considered the limitations posed by
finite memory in a simple two-hop wireline network. Inspired by this work, in \cite{VelambiITW, VelambiITA} authors investigated the information-theoretic capacity of multi-hop wireline networks with varying buffer constraints at each node. However, they only considered upper and lower bounds for the throughput. Further, the problem of finding the packet delay has been visited in the queueing theory literature on the behavior of open tandem queues that are analogous to line networks \cite{Tayfur_QT1, Tayfur_QT2}. However, this view is not applicable directly in general network topologies.

We consider wired acyclic erasure networks with ideal feedback on the links when a directed random packet routing scheme is used. Although the previous works provide some insight into the performance analysis of networks, they are limited to either infinite-buffer cases or a simple two-hop line network with limited memory. Moreover, the interplay of the throughput and latency is not considered in a small buffer regime. Our approach employs a discrete-time model to derive estimates for the buffer occupancy distribution at intermediate nodes. We then obtain analytical expressions for throughput and average packet delay in terms of the estimated buffer occupancy distributions.

The motivation behind this work is twofold: From a practical point of view, it analyzes the performance parameters of the random routing protocol in wired directed networks which is more of an interest to the network community. On the other hand, from a theoretical point of view, our work develops a framework that not only can be adapted to estimate the performance parameters of various communication schemes such as random linear coding but also can provide insight into the interplay among the buffer size, throughput and delay in general. The later is impossible to be exactly analyzed due to the exponential growth in the number of states and the complexity of the equations~\cite{VelambiITW} but our work develops an approximation to it. Further, the work establishes a lower bound on the information theoretic capacity of finite buffer networks.

This paper is organized as follows. First, we present a formal definition of the problem and the network model in Section~\ref{FB-sec1}. Next, we investigate the tools and steps for finite-buffer analysis in Section~\ref{FB-sec2}. We then obtain expressions for throughput and average delay in Section~\ref{FB-sec3}. Finally, Section~\ref{FB-sec4} presents our analytical results compared to the simulations.


\section{Problem Statement and Network Model}\label{FB-sec1}
Throughout this work, we model the network by an acyclic directed graph $\G(V,\E)$, where packets can be transmitted over a link $\e=(u,v)$ only from the node $u$ to $v$. The system is analyzed using a discrete-time model, where each node can transmit at most a single packet over a link in an epoch. The links are assumed to be unidirectional, memoryless and lossy, i.e., packets transmitted on a link $\ora{e}=(u,v)\in\ora{E}$ are lost randomly with a probability of $\ve_{\ora{e}}=\ve_{(u,v)}$. Note that the erasures are due to the quality of links (e.g., noise, interference) and do not represent packet losses due to finite buffers. Clearly, erased packets are disregarded. Moreover, the packet losses on different links are assumed to be independent. Each node $v\in V$ has a buffer size of $m_v$ packets with each packet having a fixed size. Source and destination pairs are assumed to have sufficient memory to store any data packets. The unicast information theoretic throughput between a pair of nodes is defined to be the transmission rate of information packets (in packets per epoch) between them. The delay of a packet is also defined as the time taken from the instant when the source starts transmitting a packet to the instant when the destination receives it. Note that the source node can generate innovative packets during each epoch.

Throughout this paper, node $s$ and node $d$ represent the source and destination nodes respectively. Also, for any $x\in[0,1]$, $\ol{x}\triangleq 1-x$.

\vspace{-.05 in}
\section{Understanding Finite-Buffer Analysis}
\label{FB-sec2}
\vspace{-.05 in} 
Here, we study the tools and steps that enable our framework for
analyzing finite-buffer wired erasure networks.
\vspace{-2mm}
\subsection{Communication Scheme and Buffer States}
\label{FB-sec2.1}

Thoroughly investigated in \cite{VelambiITW} for the exact analysis of a finite-buffer line network, the problem of identifying the throughput capacity is equivalent to the problem of finding the buffer occupancy distribution of the intermediate nodes as a result of ergodicity of the corresponding Markov chain. Hence, in order to approach our problem properly it is necessary to clearly define the buffer states in a manner that first of all, the buffer states construct an irreducible ergodic Markov chain and second the steady state distribution of the buffer states could be helpful to obtain expressions for the performance parameters of the network. Thus, a proper definition for the buffer states cannot be proposed unless the communication scheme is known. For example, since Random Linear Coding (RLC) is used as the communication scheme in \cite{VelambiITW}, the states are defined as the number of packets stored in the buffer of a node whose information has not yet been conveyed to the next-hop node. Analyzing RLC in a general network (which achieves the capacity of finite buffer regime) is very involved and is the subject of the future work. Instead, as a first step towards estimating the performance parameters we consider a directed random routing scheme for packets together with lossless zero-delay feedback on the links.

To be more precise, the nodes operate using the following rules, one after another.
\begin{itemize}
\item[1.] At each epoch, every node $u$ selects a random ordering of the outgoing edges and transmits the packets it houses one by one. If the packet is successfully received and stored at a neighbor, $u$ deletes the packet from its buffer and transmits the next packet (if any) on the next edge in the selected order. Else, it tries to transmit the same packet on the next (in the selected order) outgoing edge. This process is continued until all packets are transmitted or a transmission is attempted on each link. Therefore, a node with $d_o$ outgoing links transmits at most $d_o$ packets per epoch.
\item[2.] After the transmission attempts are made, the node attempts to accept the arriving packets. If more packets are received than it can store, it selects a random subset of the arriving packets whose size equals the amount of space available and stores the selected packets. Consistent with step $1$, appropriate acknowledgment messages are then sent.
\end{itemize}
Note that under such a mode of operation, at any epoch when two nodes receive two packets from a particular node, the received packets have no common information. Equivalently, no replication is performed at any node and it is possible to define a concept of state or occupancy for each node individually. The buffer state of a node is simply defined to be the number of packets it stores. It can be seen that this concept of occupancy follows a Markov chain behavior and can be studied thus.

\subsection{Approximate Markov Chain for an intermediate Node}
\label{FB-sec2.2}

\begin{figure}[htbp!]
\centering
\psfrag{a}{\hspace{-2mm}\small{$v_1$}}
\psfrag{b}{\hspace{-2mm}\small{$v_{d_i}$}}
\psfrag{c}{\small{$w_1$}}
\psfrag{d}{\small{$w_{d_o}$}}
\psfrag{e}{\small{$m_u$}}
\includegraphics[height=1.25in,width=1.5in,angle=0]{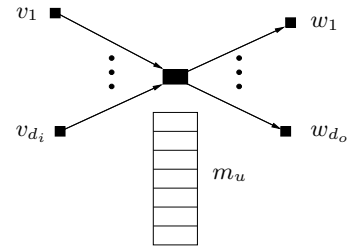}\vspace{-2mm}
\caption{A Node in a general wired network.}\label{FB-gnode}\vspace{-2mm}
\end{figure}

Consider a node $u\in V$ in a network $\ora G(V,\ora E)$ with $d_i$ incoming and $d_o$ outgoing edges and a buffer size of $m_u$ as depicted in Fig.~\ref{FB-gnode}.
Let the nodes that can send packets to $u$ be denoted by $\mc N^+(u)\triangleq\{v_i,\ldots,v_{d_i}\}$. Similarly, let the
nodes to which $u$ can send packets be denoted by $\mc N^-(u)\triangleq\{w_i,\ldots,w_{d_o}\}$. We assume that the following assumptions hold in the network regarding the arrival and departure processes.
\begin{itemize}
\item[1.] For each $k=1,\ldots,d_i$, suppose that the packets arrive on $(v_k,u)$ in a memoryless fashion with a rate of $\lambda_k$
packets/epoch \emph{i.e.,} with inter-arrival times having a geometric distribution with mean $\frac{1}{\lambda_k}$. Also, the processes on different incoming links are statistically independent.
\item[2.] At any instant, for every $k=1,\ldots,d_o$, a packet is sent on $(u,w_k)$ it is successfully received and stored at $w_k$ with a probability $\omega_k$ independent of the past and future events on the edge.
\end{itemize}
Note that this is hypothetical since in any realistic model of a network, the probability that a packet is successfully transmitted and stored at the next hop depends not only on the channel conditions, but also state of the next-hop node. Since the state of the next-hop node has dependence on its past, the probability of successful receipt can also be expected to have a dependence on its past. In fact this mode of node operation can be replaced by any other scheme that fits into the Markovian set-up of the assumptions above.

At any instant, the number of packets arriving can range from $0$ up to $d_i$ and the number of packets departing can range from $0$ to $d_o$. Hence, at each epoch, the state $n_u$ can change to any other in the set
$\{n_u-d_o,\ldots,n_u+d_i\}\cap\{0,\ldots,m_u\}$. At any epoch, the probability $a_k$ with which $k$ packets arrive and the probability $e_k$ with which $k$ packets depart are given by

\begin{figure}[ht!]
\centering
\psfrag{a}{\hspace{-1mm}\footnotesize{$0$}}
\psfrag{b}{\hspace{-1mm}\footnotesize{$1$}}
\psfrag{c}{\hspace{-1mm}\footnotesize{$2$}}
\psfrag{d}{\hspace{-1mm}\footnotesize{$3$}}
\psfrag{e}{\hspace{-1mm}\footnotesize{$4$}}
\psfrag{f}{\hspace{-1mm}\footnotesize{$5$}}
\includegraphics[width=2.25in,angle=0]{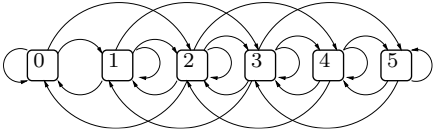}\vspace{-2mm}
\caption{The dynamics of a node $u$ with $m_u=5$ and $d_i=d_o=2$.}\label{FB-advMC}\vspace{-2mm}
\end{figure}

\vspace{-4mm}
{\footnotesize{\begin{eqnarray}
A(x)=\sum_{k=0}^{d_i}a_kx^{k}=\prod_{j=1}^{d_i}(\ol\lambda_j +\lambda_jx)\\
E(x)=\sum_{k=0}^{d_o}e_kx^{k}=\prod_{j=1}^{d_o}(\ol\omega_j +\omega_jx).
\end{eqnarray}}}
The dynamics of the number of packets stored at $u$ at the $l^{\textrm{th}}$ epoch is a Markov chain that is similar to the one depicted in Fig.~\ref{FB-advMC}. 

For all input parameters, the Markov chain can be shown to be
aperiodic, irreducible and ergodic. Therefore, it possesses a unique steady-state distribution. Letting $\Lambda=(\lambda_1,\ldots,\lambda_{d_i})$ to denote the vector of arrival rates and $\Omega=(\omega_1,\ldots,\omega_{d_i})$
to denote the vector of departure rates, the unique steady-state distribution $\vt(\cdot, \Lambda,\Omega,m_u)$ for the chain can be computed using a pair of probability transition matrices $T_{E}$ and $T_A$\footnote{For notational consistency, we can extend $e_k=0$ for $k>d_o$ and $a_k=0$ for $k>d_i$. Also, for notational convenience, we use $\vt(\cdot)$ as a short-hand $\vt(\cdot, \Lambda,\Omega,m_u)$.} that correspond to the transitions between states that are effected by the departure and arrival of packets, respectively. Note that $\vt$ is the steady-state distribution after the arriving packets are processed. These transition matrices are defined as follows.

\vspace{-2mm}
{\footnotesize{
\begin{equation}
T_E=\left[\begin{array}{cccccccc}1 & 0 & 0 & \cdots & 0& 0\\
\sum_{k=1}^{d_i} e_k & e_0 & 0 & \cdots & 0& 0\\
\sum_{k=2}^{d_i} e_k & e_1 & e_0 & \cdots & 0& 0\\
\sum_{k=3}^{d_i} e_k & e_2 & e_1 & \cdots & 0& 0\\
& & & \vdots & & &\\
\sum_{k=m_u}^{d_i} e_k & e_{m_u-1} & e_{m_u-2} & \ldots & e_1 & e_0
  \end{array}\right].\label{eqn-TEdefn}\end{equation}}}
\vspace{-2mm}
{\footnotesize{
\begin{equation}
T_A=\left[\begin{array}{cccccccc}a_0 & a_1 & a_2 & a_3 &\cdots & a_{m_u-1}& \sum_{k=m_u}^{d_i} a_k\\
0 & a_0 & a_1 & a_2 & \cdots & a_{m_u-2}& \sum_{k=m_u-1}^{d_i} a_k\\
0& 0 & a_0 & a_1 & \cdots & a_{m_u-3}& \sum_{k=m_u-2}^{d_i} a_k\\
& & &  \vdots & & &\\
0 & 0 & 0 & 0 & \cdots & a_{0}& \sum_{k=1}^{d_i} a_k\\
0 & 0 & 0 & 0 & \cdots & 0& 1\\
\end{array}\right].\label{eqn-TAdefn}\end{equation}}}

Note that, the $i,j^{\textrm{th}}$ entry in $T_E$ corresponds to the transition of the occupancy from  $i-1$ to $j-1$ with the departure of $i-j$ packets. Similarly, the $i,j^{\textrm{th}}$ entry in $T_A$ corresponds to the transition from $i-1$ to $j-1$ with the arrival of $i-j$ packets. The actual transition matrix for the Markov chain is then seen to be $T_ET_A$.
The steady-state distribution $\vt$ of the occupancy just after the arriving packets are accepted and the steady-state distribution $\vt^\dagger$ of the occupancy just after the packets have been sent but before arriving packets are
accommodated are given by
\begin{equation}
\vt T_ET_A=\vt \textrm{  and  }
\vt^\dagger T_AT_E=\vt^\dagger.
\label{EigenSSP}
\end{equation}
However, these two steady-state distributions are related by $\vt^\dagger=\vt T_E$  and $\vt=\vt^\dagger T_A$. To evaluate the rate of information on the link $(u,w_i)$, one must investigate the rule for packet departure. If at an epoch, more packets are stored than the number of links that allow successful transmission, then each link conveys a packet of information to its neighbors. However, if the occupancy $n_u$ at an epoch $l$ is smaller than the number $h$ of outgoing links that allow for transmission, each link can be assumed to equally receive $\frac{n_u}{h}$ packets on the average -- a consequence of the random selection of ordering for outgoing links. Then, the time average of the information rate on the edge $(u,w_i)$ can be seen as

\vspace{-4mm}
{\footnotesize{\begin{equation}
\begin{split}
\hspace{-14mm}I(\{(u,w_i)\},\Lambda,\Omega,m_u)=\mathop{\sum_{{H\subset\{0,\ldots, d_o\}}}}_{i\in H}\Big(\prod_{k\in H}\omega_k\Big) \times\\\Big(\prod_{k'\in H^c}\ol \omega_{k'}\Big)\Big(\sum_{j\geq |H|}\vt(j)+\sum_{j<|H|}\frac{j}{|H|}\vt(j)\Big). \label{eqn-Infor}
\end{split}
\end{equation}}}
In a similar argument, we notice that some of the arriving packets get randomly blocked if all the arriving packets cannot be stored. We can evaluate the probability with which a packet arriving on the edge $(v_i,u)$ is blocked from

\vspace{-4mm}
{\footnotesize{\begin{equation}
\begin{split}
\hspace{-14mm}
p_b(\{(v_i,u)\};\Lambda,\Omega,m_u)=\hspace{-3mm}\mathop{\sum_{{H\subset\{0,\ldots, d_i\}}}}_{i\in H} \hspace{-0.5mm}\Big(\prod_{k\in H\setminus\{i\}}\hspace{-2mm}\lambda_k\Big)\times\\  \Big(\prod_{k'\in H^c}\hspace{-1mm}\ol \lambda_{k'}\Big)\Big(\hspace{-2mm}\sum_{m_u-j<|H|}\hspace{-2mm}\frac{|H|-m_u+j}{|H|}\vt^\dagger(j) \Big). \label{eqn-block}
\end{split}
\end{equation}}}

\subsection{Iterative Estimation of the Buffer Occupancies}
\label{FB-sec2.3}

In this section, we discuss our iterative estimation technique in details based on the approximate Markov chain model introduced in section~\ref{FB-sec2.2}. Considering that blocking will introduce dependence of the packet incoming/outgoing process over each edge on its past, in order to use the results of Section~\ref{FB-sec2.2}, we have to make certain simplifying assumptions on the blocking phenomenon. We model the blocking on every edge $\ora e =(u,v)$ of the network as follows.
\begin{itemize}
\item Every packet that arrives at $v$ successfully (without getting erased) is blocked in a memoryless fashion with probability $q_{uv}$. Also, at any epoch, the blocking of packets on any subset of incoming edges of $v$ is assumed to
be independent of one another.
\end{itemize}
Under the above assumption, the blocking process and hence the departure process on every link of the network is modeled as a memoryless process. Since each packet arriving on an edge $\ora e=(u,v)$ is blocked with a probability of $q_{uv}$, a packet arriving on $\ora e$ is accepted only if both the channel allows the packet and the node accepts it. Therefore, the effective departure rate on the edge $(u,v)$ is seen to be $\ol\ve_{uv}\ol q_{uv}$. Assuming that the node operates in the mode described in Section~\ref{FB-sec2.1}, we can use (\ref{eqn-Infor}) and (\ref{eqn-block}) to identify both the rate of information flow and the blocking probabilities on every edge of the network. Thus, the problem reduces to finding a solution $(\varrho_{uv},q_{uv})_{(u,v)\in\ora E}$ that satisfies the following system of non-linear equations for each $(u,v)\in\ora E$.

\vspace{-3mm}
 {\allowdisplaybreaks
\footnotesize{\begin{eqnarray*}
\varrho_{uv}\hspace{-3mm}&=&\hspace{-3mm}\left\{\hspace{-2mm}\begin{array}{lc} \ol\ve_{uv} & {u=s}\\
\frac{I(\{(u,v)\},(\varrho_{wu})_{w\in\mc N^+(u)},(\ol\ve_{uu'}\ol q_{uu'})_{u'\in\mc N^-(u)},m_u)}{\ol q_{uv}} &{u\neq s}\end{array}\right., \label{eqn-sys1}\\
q_{uv}\hspace{-3mm}&=&\hspace{-3mm} \left\{\hspace{-2mm}\begin{array}{lc} p_b(\{(u,v)\};(\varrho_{wv})_{w\in\mc N^+(v)},(\ol\ve_{vv'}\ol q_{vv'})_{v'\in\mc N^-(v)},m_v) & v\neq d\\
0 & v=d\end{array}\right. \label{eqn-sys2}.
\end{eqnarray*}}}
Note that in the above equations $\varrho_{uv}$ represents the fraction of time at which packets will be delivered to $v$. However, the actual rate of information flow is equal to $\rho_{uv}=\ol q_{uv}\varrho_{uv}$.

Since the above set of equations are an approximation to the actual dynamics, it is not clear as to whether there even exists a solution to the above system. However, the proof of existence and uniqueness of the solution is detailed in~\cite{Vellambi2010} for the case of line networks.

Finally, the solution to the system of equations can be found by identifying the limit of the sequence defined by the following iterative procedure\footnote{In practice, the number of iterations $L$ which suffice to converge to the solution within reasonable accuracy depends on the structure of the network. Alternatively, one may use $|{\vt}^{(i+1)}-{\vt}^{(i)}|+|{\vt^\dagger}^{(i+1)}-{\vt^\dagger}^{(i)}|<\epsilon$ for the convergence criteria.}.
\begin{itemize}
\item[1.] Set $i=1$ and for each edge $(u,v)\in \ora E$, set $q_{uv}=0$ and $\varrho_{uv}^{(1)}=\left\{\begin{array}{cc} 0 &{u\neq s} \\ \ol\ve_{uv} & {u=s} \end{array}\right.
$.
\item[2.] Compute $\varrho_{uv}^{(i+1)},q_{uv}^{(i+1)}$ by using $\varrho_{uv}^{(i)},q_{uv}^{(i)}$ on the right-hand side of the above system of nonlinear equations and increment $i$ by 1.
\item[3.] If $i<L+1$, perform step 2.
\end{itemize}

\section{Estimation of the Throughput and Average Packet Delay}
\label{FB-sec3}

In this section, we exploit the results of the iterative estimation method for buffer occupancy distributions and obtain analytical expressions for throughput and average delay.

Since the routing scheme is such that information is not replicated at any node, the estimate of the total information that arrives at the destination is the sum total of the information rate arriving on each of its incoming edges. Hence,
\begin{equation}
\hat{\mcc}(s,d,\ora G)=\sum_{v\in\mc N^+(d)} \varrho_{vd}^*(1-q_{vd}^*)=\sum_{v\in\mc N^+(d)} \varrho_{vd}^*,
\end{equation}
where we let $(\varrho^*_{uv}, q_{uv}^*)$ to be either the component-wise limit of the sequence $\{\varrho_{uv}^{(i)},q_{uv}^{(i)}\}_{i\in\mathbbm{N}}$ when $L=\infty$, or $(\varrho_{uv}^{(L)},q_{uv}^{(L)})$ when
$L<\infty$. Additionally, by the conservation of information flow, the above estimate can be obtained by computing the rate
of flow of information through any cut $F$ using the following.
\begin{equation}
\hat{\mcc}(s,d,\ora G)=\sum_{uv\in F} \varrho_{uv}^*(1-q_{uv}^*).
\end{equation}

 As it is defined in Section~\ref{FB-sec2.1}, the operation scheme is chosen to have feedback on all the links and we treat packets in a First-Come First-Serve (FCFS) fashion at the buffers. Also, the absence of directed cycles allows us to assign an order ${v_1,v_2,\ldots,v_n}$ to all the nodes of the network in a manner that we have $i<j$ for every link $(v_i,v_j)\in\ora{E}$.

In order to estimate the average delay that a packet experiences in the network, one can proceed in a recursive fashion. The average delay that an arriving packet (at time $l$) at node $u\in V$ experiences depends on the buffer occupancy of the node $u$ and its outgoing links. For example, suppose at epoch $l$ (packet arrival time), node $u$ has already $k$ packets where $k \leq m_u-1$. Then, the arriving packet has to wait for the first $k$ packets to leave node $u$ before it can be transmitted. We define $\mc D_u(k)$ as the average time it takes from the instant that node $u$ receives a packet given that it has already stored $k$ packets, until the time that the destination node receives that packet. We compute the average delay function $\mc D_u(.)$ for all the intermediate nodes $u\in V$ using the following proposition. 

\vspace{2mm}{
\begin{prop}
Let $r_{uv}=\ol \ve_{uv} \ol q_{uv}$ be the average packet transfer rate on link $(u,v)\in\ora{E}$ and $r_u^-$ be the sum of the rates on all outgoing edges (\emph{i.e.}, $r_u^-=\sum_{v\in \mc N^-(u)} {r_{uv}}$). Also, let $\pi_v (j)$ ($j=0,1,\ldots,m_v-1$) be the steady state probability of the buffer of node $v\in V$ storing already $j$ packets right before a new packet arrives and is stored in the buffer. For every intermediate node $u\in V$, given the average delay functions of all its next-hop neighbors ($\mc D_v(j)$ for all $v\in \mc N^-(u)$ and $j=0,1,\ldots,m_v-1$), $\mc D_u(.)$ can be obtained by

\vspace{-1mm}
{\footnotesize{
\begin{equation}
\mc D_u(k) = \frac{k+1}{r_u^-} + \sum_{w\in \mc N^-(u)} { \frac{r_{uw}}{r_u^-} \Big( \sum_{j=0}^{m_{w}-1} {\pi_w (j) \mc D_w (j)} \Big) }
\label{Delay-Eqn}
\end{equation}}}
for $k=0,1,\ldots,m_u-1$.
\label{ADprop}
\end{prop}}

\vspace{1mm}{
\begin{proof}
Due to space limitation, we just provide a brief sketch of the proof.
Equation (\ref{Delay-Eqn}) is consisted of two terms:
\begin{itemize}
\item[1.] The first term represents the average time it takes for a total of $k+1$ packets (counting our selected packet) to leave node $u$ successfully.

\item[2.] The second term relates to the average delay due to the rest of the network. The probability of conveying a packet from node $u$ to node $v$ can be estimated by $\frac{r_{uv}}{r_u^-}$. An arriving packet at node $v$ finds its buffer already occupied by $j$ packets with probability $\pi_v (j)$. Thus, the packet will experience an average delay of $\mc D_v(j)$ from this node to the destination. Hence, the average packet delay from node $v$ to the destination is equal to $\sum_{j=0}^{m_{w}-1} {\pi_w (j) \mc D_w (j)}$.

\end{itemize}
\end{proof}}

It is easy to see that $\pi_v (j)$ can be calculated using
\begin{equation}
\pi_v (j)=\left\{\begin{array}{ll} \frac{\vt_v^\dagger(j)}{1-\vt_v^\dagger(m_v)} & j=1,2,\ldots,m_v-1\\ 0 & j=m_v
\end{array}\right.
\end{equation}
To obtain the average packet delay from the source to the destination, the average delay function $\mc D_u(.)$ is computed for all the nodes in the reverse order\footnote{Note that we have $\mc D_d(k)=0$ for every k where $d$ denotes the destination node.} (\emph{i.e.},$\{v_n,\ldots, v_2, v_1\}$). Then, the total average packet delay ($\mc D_s(0)$) is computed by applying Proposition~\ref{ADprop} to the source node.


\section{Simulation Results}
\label{FB-sec4}
\vspace{-.05 in} 

In this section, we present the results of actual network simulations in comparison with our analysis and will show that our framework gives accurate estimates of buffer occupancy distributions as well as throughput and average delay.

\begin{figure}[htbp!]
\centering
\includegraphics[width=1.25in,angle=-90]{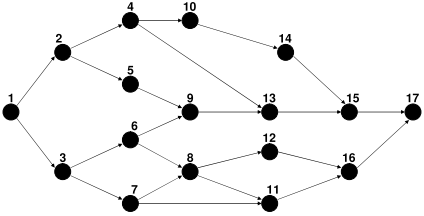}
\caption{A general wired acyclic directed network chosen for simulation).}\label{FB-GNet1}
\end{figure}

We consider the network shown in Fig.~\ref{FB-GNet1} to compare the results of the simulation and inferences.
In this network, all the edges have $\ve=0.5$ (erasure probability) except the edges $\{(1,2), (1,3), (15,17), (16,17)\}$ for which $\ve=0.05$.
All the intermediate nodes are assumed to have the same buffer size. In order to measure the exact performance parameters of this network, millions of packets are sent from the source (Node $1$) to the destination (Node $17$).
\begin{figure}[ht!]
\centering
\psfrag{xaxis}{\small{\hspace{-3mm}Buffer occupancy (packets)}}
\psfrag{yaxis}{\small{Probability}}
\psfrag{Node 3}{Node 3}
\psfrag{Node 4}{Node 4}
\psfrag{Node 11}{Node 11}
\psfrag{Node 15}{Node 15}
\psfrag{Simulation}{\tiny{Simulation}}
\psfrag{Analysis}{\tiny{Analysis}}
\includegraphics[width=3.75in,height=2.25in,angle=0]{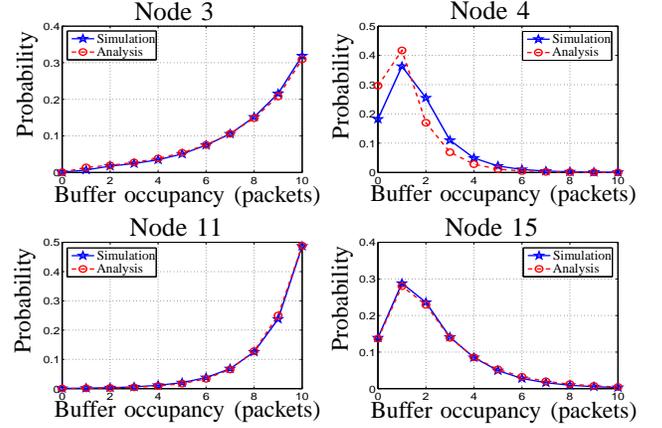}
\caption{buffer occupancy distributions for nodes $3$, $4$, $11$ and $15$.}\label{Net1_BuffOcc}
\end{figure}
\begin{figure}[ht!]
\centering
\psfrag{YaxisT}{\small{\hspace{-3mm}Throughput}}
\psfrag{XaxisT}{\small{Buffer size $m$}}
\psfrag{YaxisB}{\small{\hspace{-3mm}Average delay}}
\psfrag{XaxisB}{\small{Buffer size $m$}}
\psfrag{Simulation11}{\footnotesize{Simulation}}
\psfrag{Analysis}{\footnotesize{Analysis}}
\includegraphics[width=3.75in,height=2.25in,angle=0]{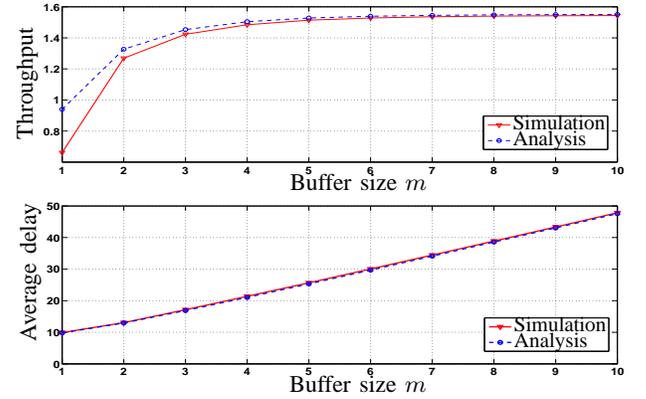}
\caption{Performance parameters of the network for different buffer sizes.}\label{Net1_Results}
\end{figure}
Fig.~\ref{Net1_BuffOcc} presents a comparison between the actual buffer occupancy distributions and our iterative estimates for four of the nodes in the network of Fig.~\ref{FB-GNet1}. Also, Fig.~\ref{Net1_Results} presents the variations of the actual throughput and average packet delay and our analytical results versus the buffer size. Note that, the throughput is presented in \emph{packets/epoch} and average packet delay is presented in \emph{epochs}. As it can be observed, our estimation is very close to the actual simulation results.

\bibliographystyle{ieeetr}
\bibliography{ITWRef}
\end{document}